\documentclass[prl,twocolumn,showpacs,preprintnumbers,amsmath,amssymb]{revtex4}
\usepackage{graphicx}
\usepackage{dcolumn}
\usepackage{bm}
\usepackage{color}

\begin{document}
\title{Anisotropic Ru$^{3+}$ 4$d^5$ magnetism in the $\alpha$-RuCl$_3$ honeycomb system: susceptibility, specific heat and Zero field NMR}
\author{M. Majumder$^{1}$}
\email{mayukh.cu@gmail.com}
\author{M. Schmidt$^1$}
\author{H. Rosner$^1$}
\author{A.~A. Tsirlin$^{1,2}$}
\author{H. Yasuoka$^1$}
\author{M. Baenitz$^1$}

\affiliation{$^1$Max Plank Institute for Chemical Physics of Solids, 01187 Dresden, Germany}
\affiliation{$^2$National Institute of Chemical Physics and Biophysics, Tallinn, Estonia}

\date{\today}

\begin{abstract}
Hexagonal $\alpha$-Ru trichloride single crystals exhibit a strong magnetic anisotropy and we show that upon applying fields up to 14 T in the honeycomb plane the successive magnetic order at $T_1$ = 14 K and $T_2$ = 8 K could be completely suppressed whereas in the perpendicular direction the magnetic order is robust. Furthermore the field dependence of $\chi$(T) implies coexisting ferro- and antiferromagnetic exchange between in-plane components of $Ru^{3+}$-spins, whereas for out-of-plane components a strong antiferromagnetic exchange becomes evident. $^{101}$Ru zero-field nuclear magnetic resonance in the ordered state evidence a complex (probably non coplanar chiral) long-range magnetic structure. The large orbital moment on Ru$^{3+}$ is found in density-functional calculations.

\end{abstract}
\maketitle

Low dimensional 4$d$- and 5$d$-magnets show a wide variety of magnetic ground states due to crystal electric field (CEF) splitting in combination with a strong spin-orbit coupling (SOC). Especially the $5d^5$- iridate compounds earned great attention because of the predicted topological Mott insulating state due to the strong SOC and the Coulomb correlation\cite{Pesin10}. Furthermore the strong SOC favors the asymmetric Dzyaloshinskii-Moriya (DM) interaction that often results in chiral spin arrangements in the ordered phases\cite{Dzyaloshinskii64,Moriya60}. In addition, for spin-1/2 systems geometrical frustration of the magnetic exchange interactions frequently leads to a quantum spin-liquid ground state\cite{Balents10}. Among 4$d$- and 5$d$- systems, the Heisenberg-Kitaev model (HKM) was established to describe the competing bond-dependent magnetic exchange interactions in the honeycomb type of lattice structures\cite{Kitaev}. Prominent examples are the 2-1-3 iridates (Li$_2$IrO$_3$, Na$_2$IrO$_3$) where the magnetism is associated to the 5$d^5$ electrons on the Ir$^{4+}$ ions. According to the HKM, the phase diagram provides a transition from a conventional Neel-type of antiferromagnetic (AFM) order to a AFM stripy- (or zigzag-) type of order towards a pure quantum spin liquid (QSL) phase as a function of control parameter\cite{Challoupka10}. Indeed Na$_2$IrO$_3$ shows an AFM order of zigzag-type at T = 15 K, whereas the Li$_2$IrO$_3$ system is more close to the QSL regime and a non coplanar spiral order is discussed\cite{Singh12}.

In order to search for new 4$d$- or 5$d$- model system with the honeycomb lattice arrangement as a platform of HKM $\alpha$-RuCl$_3$ turns out to be an excellent candidate because the low spin 3+ state of Ru ($4d^5$) is equivalent to the low spin 4+ state of Ir ($5d^5$). However, low-temperature magnetic properties of $\alpha$-RuCl$_3$ were not studied in detail and, so far, on powders only. Recently, Plumb and co-workers have shown in a spectroscopic experiment that $\alpha$-RuCl$_3$ is a magnetic insulator due to sizable Coulomb correlations accompanied by the spin-orbit coupling\cite{Plumb14}.

In this Rapid Communication, we report detailed studies on the magnetic anisotropy by magnetization and specific heat on single crystals over a wide temperature and field range. Furthermore, we applied $^{99,101}$Ru zero-field nuclear magnetic resonance as a local and "on-site" probe for the ordered ground state. The system has a strong magnetic anisotropy and reveals an overall ferromagnetic exchange when field is applied in the plane ($H\perp c$) and an AFM one when the field is applied perpendicular to the plane ($H\parallel c$). Moreover, two phase transitions at $T_1$ = 14 K and $T_2$ = 8 K have been detected. We show further that upon applying fields up to 14 T in the plane, the complex magnetic order could be completely suppressed, whereas in the perpendicular direction the magnetic order is robust up to 14 Tesla.

Single crystals of $\alpha$-RuCl$_3$ were grown by chemical transport reaction, starting from a microcrystalline powders of pre-reacted materials, which were obtained by the reaction of the elements. The transport experiment was carried out from microcrystalline sample of $\alpha$-RuCl$_3$ in an evacuated quartz tube in a temperature gradient from 730$^\circ$C (source) to 660$^\circ$C (sink). Chlorine (2 mg/ml) was used as transport agent. Selected crystals were analyzed by EDXS (which provides a typical stoichiometry of Ru:Cl = 1:3.09), chemical and thermal analysis, and X-ray diffraction. $\alpha$-Ru trichloride forms a hexagonal structure (P3$_1$12) with $a$=5.97 ${\AA}$ and $c$=17.2 ${\AA}$\cite{Fletcher67}. The structure hosts one Ru site and three different Cl sites. The crystals have lateral dimensions of a few mm, whereas their thickness is less than 0.1 mm. Magnetization measurements were performed with a commercial SQUID VSM (vibrating sample magnetometer) from Quantum Design (QD) and using the PPMS magnetometer option. Specific heat measurements are conducted at a commercial PPMS (Physical Property Measurement System) from Quantum Design by use of a modified sample holder for "vertical" (field in-plane) specific heat measurements. Magnetization measurements in pulsed fields up to 50 T are performed on powder samples at the high field research center (HLD) at Helmholtz-Zentrum Dresden-Rossendorf. All magnetic measurements are performed in the ZFC (zero field cooled) modes. Between field cycles the sample was always heated to 200 K and then the field was ramped to zero. Magnetization and specific heat measurements are performed on a single crystal ($\sim$ 1 mg), whereas zero field NMR was done on a stack of single crystals ($\sim$ 90 mg). NMR measurements are carried out by applying conventional pulsed NMR. The zero-field NMR spectra have been obtained by sum of the spin echo time FFT (Fast Fourier Transform) spectra measured typically 0.2 MHz step. In general, the zero field spectra originate from the anisotropic hyperfine field transferred from the Ru-4$d^5$ moments.

\begin{figure}[h]
{\centering {\includegraphics[width=0.5\textwidth]{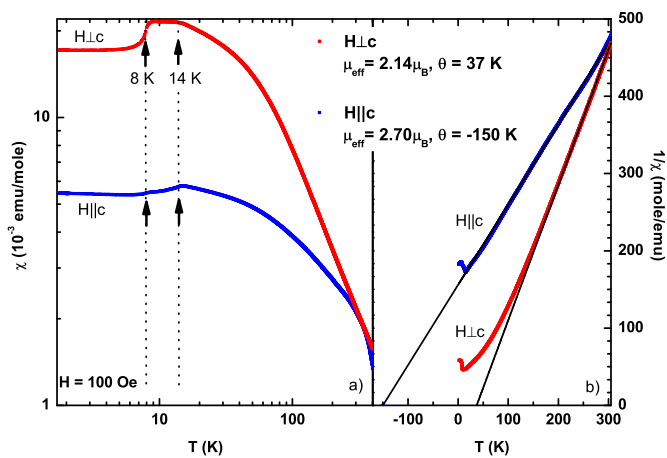}}\par} \caption{(Color online) (a): Temperature dependence of the magnetic susceptibility ($\chi_\parallel$ and $\chi_\perp$) of $\alpha$-RuCl$_3$ measured in H=100 Oe, (b): Inverse susceptibilities and Curie Weiss fit (straight lines).} \label{structure}
\end{figure}

Figure 1(a) shows the susceptibility as a function of temperature for the two directions (in-plane and perpendicular to the plane) in $\alpha$-RuCl$_3$. For H$\|$c there are clear signatures of two phase transitions at 14 and 8 K, whereas for H$\bot$c between 14 and 8 K a plateau-like behavior is found, followed by a sharp decrease of the susceptibility. This decrease points towards a sort of spin compensation (eg., an antiferromagnetic transition or an effective dimerization in the $ab$ plane). Below 300 K, the in-plane susceptibility $\chi_\perp$ is larger than the out-of-plane component $\chi_\parallel$ ($\chi_\perp/\chi_\parallel$ $\approx$ 2.5 at 60 K). The inverse susceptibility is shown in figure 1(b). The Curie Weiss fit of the inverse susceptibility clearly yields dissimilar effective coupling constants for the two field directions: i) for H$\bot$c a Curie Weiss temperature of + 37 K evidences an effective ferromagnetic exchange ($J/K_B$=37 K) and an effective moment of 2.14 $\mu_B$/Ru and ii) for H$\|$c a Curie Weiss temperature of -150 K which evidences a strong antiferromagnetic effective exchange and an effective moment of  2.7 $\mu_B$. So far no single crystal data for magnetization were available in the literature. From powder results a positive Curie Weiss temperature $\theta$ = 23-40 K was determined \cite{Kobayashi92} and the moments calculated are around 2.3 $\mu_B$. This is approximately what we found for the in-plane contribution in our single crystal and it is most likely that the powder average is dominated by this in-plane contribution. Nonetheless, our single crystal studies clearly reveal the anisotropic nature of the magnetic exchange and give clear evidence for an effective antiferromagnetic exchange interaction with a larger effective moment of 2.7 $\mu_B$ when field is applied along the $c$ direction. This is well above the spin-only value for the $S=1/2$ low spin configuration of Ru (1.73 $\mu_B$) which points towards a prominent SOC. This is supported by our field-dependent susceptibility studies shown in figure 2(a). In the $c$ direction, the susceptibility and the step-like transitions at 14 K and 8 K remain unaffected in fields up to 7 T. In contrast to that, the in-plane susceptibility is strongly affected by magnetic fields for T$<$50 K. Upon applying fields upto 14 T the phase transitions are shifted towards lower temperatures and the susceptibility is enhanced. This is not expected because of the effective ferromagnetic coupling evidence from the Curie Weiss fit. The reduction of $\chi_\perp$ with field evidences admixed antiferromagnetic correlations which supports the scenario of dissimilar exchange interaction (Heisenberg type versus Kitaev type) being present in the honeycomb layer.

\begin{figure}[h]
{\centering {\includegraphics[width=0.5\textwidth]{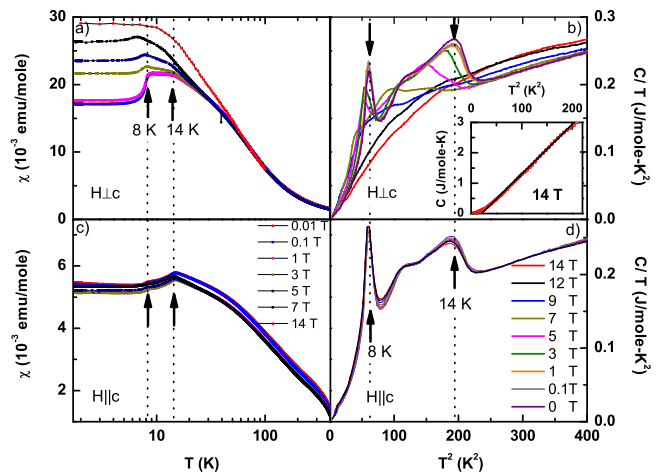}}\par} \caption{(Color online) $\chi$ for $H\perp c$ and $H\parallel c$ at different magnetic fields, (C/T) versus $T^2$ at different magnetic fields for $H\perp c$ and $H\parallel c$.} \label{structure}
\end{figure}

To study the complex phase transitions and the effect of the magnetic field for both directions, we conducted specific heat measurements (figure2). In figure 2 (b), the quotient of specific heat and temperature is plotted as a function of $T^2$ between 2K and 20 K in fields up to 14 T and for the H$\bot$c configuration. At zero field, the two phase transitions at 14 K and 8 K could be clearly identified. The high temperature transition is somewhat broader than the low-temperature transition. Upon applying magnetic fields up to 14 T, the complex transition is monotonously suppressed which is consistent with our findings in the susceptibility. At highest fields of 14 T, the onset of the low-T transition is at about 3 K. Above this transition, the $C/T$ curve remains smooth and shows a nice $C \sim T^2$ (inset of Figure 2(b)) power law between 5 K and 14 K, which might indicate a spin-liquid-like behavior. In contrast to that, there is no such strong field dependence for the H$\|$c configuration. Here, both phase transitions at 14 K and 8 K remain unaffected by the field. Assuming a negligible phonon contribution, we find an entropy of $S\approx0.5R\ln2$ below 20 K for $H=$ 0 T. So far we do not have a proper phonon reference to do a more quantitative specific heat analysis.

As a microscopic probe of the anisotropic magnetic order, we performed $^{99,101}$Ru zero-field NMR on the $\alpha$-RuCl$_3$ single crystal. Due to the presence of two isotopes with the higher ($>$1/2) spin (5/2 for $^{99}$Ru and 5/2 for $^{101}$Ru) a complex zero-field NMR spectrum is expected. Early $^{99}$Ru Moessbauer measurement on the $\alpha$-RuCl$_3$ powder show the absence of sizable quadrupolar interaction and provide the powder-averaged hyperfine field at the Ru- site of about 20.9 T\cite{Kobayashi92}. The value of the hyperfine field is in good agreement with the prediction of Watson and Freeman of about 20 T per unpaired 4$d$ electron (and 11 T per unpaired 3$d$ electron)\cite{Freeman65}. For the Ru$^{3+}$ 4$d^5$ state in the low-spin configuration (LSC) a hyperfine field of 20 T is expected whereas for the Ru$^{4+}$ 4$d^4$ state in LSC in the perovskite-type SrRuO$_3$ a hyperfine field of about 40 T is expected and experimentally confirmed\cite{Gibb71,Mukuda99}.

\begin{figure}[h]
{\centering {\includegraphics[width=0.5\textwidth]{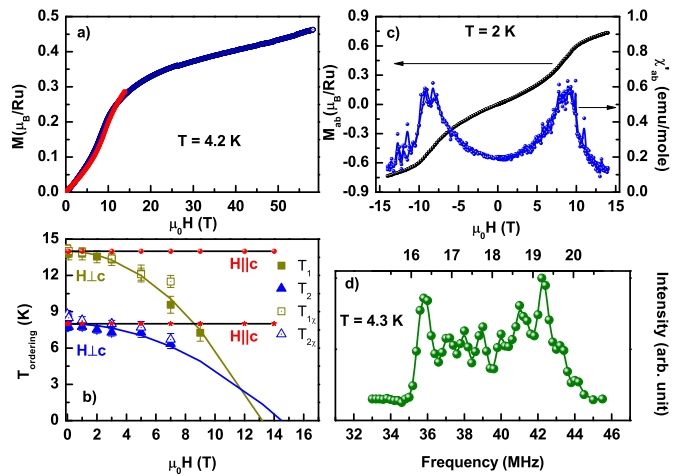}}\par} \caption{(Color online) (a) Magnetization versus field (open symbols and solid line represent data from PPMS -pulse field measurements respectively), (b) Anisotropic field tuning in RuCl$_3$ (closed symbols: heat capacity, open symbols: susceptibility). Curved solid lines correspond to a fit with T$_{ordering} = T_0 (1-(\frac{H}{H_C})^2)$ with $H_C$ = 13.2 T for the 14 K transition and $H_C$ = 14.5 T for the 8 K transition. (c)$M(H)$ and $\chi_{ac}(H)$ for $H\perp c$, (d) Zero field $^{101}$Ru NMR spectra at 4.3 K.} \label{structure}
\end{figure}

From the hyperfine field of $H_{hf}$ = 20.9 T we could calculate the NMR frequency by the gyromagnetic ratio $^{101}\gamma$ = 2.193 MHz/T which leads to an average NMR frequency of about 45.83 MHz. Figure 3(d) shows the $^{101}$Ru zero-field NMR frequency spectra at 4.3 K. The 4.3 K spectra is located somewhat lower in frequency than expected and, moreover, the line exhibits a pronounced broadening of about 10 MHz which corresponds to 4.56 T. The observed broadening of the zero field $^{101}$Ru NMR line around the average hyperfine field of about 17.8 T points toward a complex type of magnetic order. The main contribution is the on site contribution (due to core polarization) from the Ru$^{3+}$ ion to the nucleus whereas the modulation of the field ("double horn spectra") is probably related to the anisotropic hyperfine interaction with nearest and next nearest (in - plane-) neighbors in a complex moment arrangement. This could be either the planar "zig-zag" type or the "stripy" type of order predicted by the HKM or even a non planar (spiral) type found in frustrated Heisenberg chains (e.g. LiCu$_2$O$_2$\cite{Gippius04}) or discussed recently for the $5d^5$ system Li$_2$IrO$_3$\cite{Reuther}. For the isostructural high spin homologues FeCl$_3$ ($3d^5$) and CrCl$_3$ ($3d^3$) such a complex modulated line was not reported in Fe- (Cr) zero field NMR results\cite{Kang14,Narath61}.

To probe the ordered moments, we performed magnetization measurement in pulse field (H $\leq$ 60 T) on a powder specimen (Fig. 3(a)). To calibrate the high-field data, dc magnetization measurement performed in static fields up to 14 T (Figure 3(a)) were used. Up to 60 T there is no full saturation of the ordered moment in the powder specimen but from the changeover in curvature one might speculate about an in plane saturation field of nearly 15 T. To proof that we conducted $M(H)$ measurements on a single crystal with fields up to 14 T applied in the plane (Fig. 3(c)). Indeed the 14 T magnetization is close to 1$\mu_B$ which suggests that we are near full in plane saturation. Furthermore the in plane ac susceptibility shows a clear peak at about 8 T which is indicative for a spin flip due to the quenching of the in-plane afm exchange contribution in agreement with the monotonous suppression of the 8 K transition (Fig 2a).

Finally, we briefly discuss microscopic features of $\alpha$-RuCl$_3$\cite{note}. Plumb~\textit{et al.}\cite{Plumb14} have shown that ruthenium ions are in the 3+ state, and the formation of the insulating state is driven by both spin-orbit coupling (SO) and on-site electronic correlations ($U$). To explore the electronic structure of $\alpha$-RuCl$_3$, we performed full-relativistic density-functional calculations using the \texttt{FPLO} code\cite{fplo} with the basis set of local orbitals and Perdew-Wang exchange-correlation potential (LDA)\cite{pw92}.

\begin{figure}
\includegraphics{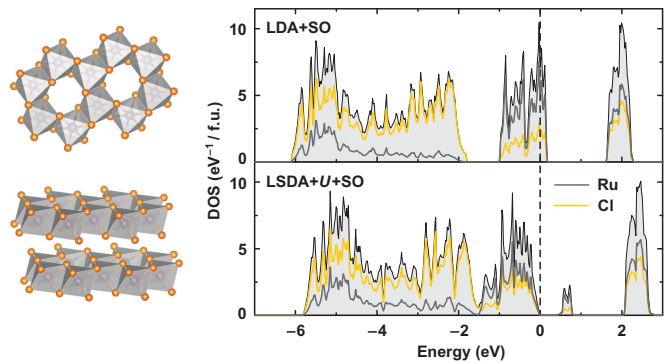}
\caption{\label{fig:dos}
(Color online) Left panel: crystal structure of $\alpha$-RuCl$_3$ built of honeycomb Ru--Cl layers that are held together by weak van-der-Waals forces. Right panel: electronic density of states (DOS) for $\alpha$-RuCl$_3$ calculated within LDA+SO (top) and LSDA+$U$+SO (bottom, total DOS for both spin directions is shown). The Fermi level is at zero energy. Note that the spin-orbit coupling does not split the broad complex of $t_{2g}$ bands crossing the Fermi level. The gap is formed only in LSDA+$U$+SO, where correlation effects are accounted for in a mean-field fashion.
}
\end{figure}

The LDA+SO band structure is shown in Fig.~\ref{fig:dos}, top right. It is very similar to the scalar-relativistic band structure obtained in LDA (without SO, not shown). The band structure is metallic, with states at the Fermi level dominated by Ru $4d$ orbitals. The fraction of Cl $3p$ states is about 34\% (27\% in the $t_{2g}$ bands and 44\% in the $e_g$ bands), which is comparable to 35\% of oxygen $2p$ states in honeycomb iridates. Therefore, $\alpha$-RuCl$_3$ and Ir$^{4+}$ compounds feature same degree of metal-ligand covalency. Two well-separated band complexes centered at around $-0.5$\,eV and 2.0\,eV belong to the $t_{2g}$ and $e_g$ states of Ru, respectively, but no further splitting of the $t_{2g}$ levels can be observed. In contrast to iridates,\cite{mazin2012} neither crystal-field effects nor the spin-orbit coupling split the broad $t_{2g}$ complex into narrow bands that would be then easily split by even weak electronic correlations. We quantified orbital energies by fitting the LDA+SO band structure with Wannier functions and found $\varepsilon_{xy}=-0.33$\,eV, $\varepsilon_{yz}=-0.34$\,eV, $\varepsilon_{xz}=-0.36$\,eV, $\varepsilon_{3z^2-r^2}=2.00$\,eV, and $\varepsilon_{x^2-y^2}=1.89$\,eV, where $x$, $y$, and $z$ axes are directed along the Ru--Cl bonds within the RuCl$_6$ octahedra, and the $LS$-basis is used. Similar energies of the $xy$-, $yz$-, and $xz$-orbitals reflect marginal distortion of the octahedra and suggest minor importance of crystal-field effects.

In order to reproduce the effect of electronic correlations, we performed LSDA+$U$+SO calculations assuming the ferromagnetic spin arrangement in $\alpha$-RuCl$_3$. This simplest possible spin configuration enables us to focus on the local physics of one Ru$^{3+}$ ion. Following Ref.~\onlinecite{Plumb14}, we used the on-site Coulomb repulsion $U=1.5$\,eV and Hund's exchange $J=0.3$\,eV. This leads to the insulating state with an energy gap of about 0.4\,eV in agreement with the experimental optical gap\cite{Plumb14}. Five out of six $t_{2g}$ states nearly merge into the Cl $3p$ band, whereas the remaining empty band is slightly above the Fermi level.

The spin and orbital moments on Ru$^{3+}$ in LSDA+$U$+SO are about 0.51 $\mu_B$ and 0.40 $\mu_B$, respectively, when spins are in the $ab$ plane. For spins directed along the $c$ axis a much lower spin moment of only 0.16 $\mu_B$ and a quite high orbital moment of 0.38 $\mu_B$ are found. The formation of the $j_{\rm eff}=\frac12$ state is reminiscent of iridates\cite{kim08}. Given one unpaired electron per Ru$^{3+}$ ion, the spin moment is dramatically reduced with respect to its free-ion value of 1 $\mu_B$, and a huge orbital moment is formed. This is in agreement with the experimental effective magnetic moments of 2.14 $\mu_B$ and 2.70 $\mu_B$ for $H\!\perp\!c$ and $H\|c$, respectively, that imply an effective spin-$\frac12$ behavior but with the large deviation from the spin-only value due to the orbital moment. Considering $\mu_{\rm orb}=0.40$\,$\mu_B$, we estimate $g=2.8$ and $\mu_{\rm eff}=2.42 \mu_B$ in reasonable agreement with the experiment. However, the directional anisotropy of the effective moment (different values depending on the field direction) is not fully reproduced and requires further analysis.

In contrast to iridates, the spin-orbit coupling neither splits the $t_{2g}$ manifold nor opens the band gap in $\alpha$-RuCl$_3$ when Coulomb correlations are not taken into account. The combination of the spin-orbit coupling and Coulomb correlations within LSDA+$U$+SO calculations is sufficient to reproduce the electronic structure of the material\cite{Plumb14} and pinpoints a sizable orbital moment that is likely responsible for the strong magnetic anisotropy. We have studied this anisotropy on single crystals and demonstrated that Curie-Weiss temperatures, which measure the effective magnetic exchange, are not only numerically different for different field directions but also differ in sign being positive and strongly field dependent for the field in-plane and negative and robust for the field out-of-plane. This strongly suggests that $\alpha$-RuCl$_3$ is a platform for the HKM. Details of this physics require further investigation, though, and a careful evaluation of individual exchange parameters\cite{Kubota15}. From a phenomenological viewpoint, $\alpha$-RuCl$_3$ reveals very peculiar physics with two transitions at 14 K and 8 K that are also strongly anisotropic. The $^{101}$Ru zero-field NMR line point towards a complex long-range magnetic order. Single-crystal neutron diffraction studies are required to obtain the magnetic structure in the ordered phase\cite{Sears}.

Acknowledgements: The authors thank H. Rave and C. Klausnitzer for technical support on the heat capacity measurements and A. Henschel for support on the sample preparation. Furthermore we acknowledge the support of the high field research centre (HLD) at the Helmholz-Zentrum Dresden-Rossendorf (HZR).


\begin{thebibliography}{50}
\bibitem{Pesin10} Dmytro Pesin and Leon Balents, Nature Phys., \textbf{6}, 376 (2010).
\bibitem{Dzyaloshinskii64} I. E. Dzyaloshinskii, Sov. Phys. JETP, \textbf{19}, 960 (1964).
\bibitem{Moriya60} Toru Moriya. Phys. Rev. \textbf{120}, 91 (1960).
\bibitem{Balents10} L. Balents, Nature, \textbf{464}, 199 (2010).
\bibitem{Kitaev} Alexei Kitaev. Ann. Phys. (N. Y)., 321(1):2-111, (2006).
\bibitem{Challoupka10} J. Chaloupka, G. Jackeli, G. Khaliullin, Phys. Rev. Lett. \textbf{105}, 027204 (2010), J. Chaloupka, G. Jackeli, G. Khaliullin, Phys. Rev. Lett. \textbf{110}, 097204 (2013).
\bibitem{Singh12} Y. Singh, S. Manni, J. Reuther, T. Berlijn, R. Thomale, W. Ku, S. Trebst, and P. Gegenwart, Phys. Rev. Lett. \textbf{108}, 127203 (2012).
\bibitem{Plumb14} K. W. Plumb, J. P. Clancy, L. J. Sandilands, V. V. Shankar, Y. F. Hu, K. S. Burch, Hae-Young Kee, and Young-June Kim, Phys. Rev. B \textbf{90}, 041112 (2014).
\bibitem{Fletcher67} J. M. Fletcher, W. E. Garnder, A. C. Fox, G. Topping, J. Chem. Soc. (A), 1038 (1967).
\bibitem{Kobayashi92} Y. Kobayashi, T. Okada, K. Asai, M. Katada, H. Sano, and F. Ambe, Inor. Chem. \textbf{31}, 4570-4574 (1992).
\bibitem{Freeman65} A. J. Freeman, R. E. Watson in "Magnetism" by G. T. Rado, H. Suhe, NY, \textbf{2A}, 259 (1965).
\bibitem{Gibb71} T. C. Gibb, R. Greatrex, N. N. Greenwood, and P. Kaspi, Chem. Comm, 319 (1971).
\bibitem{Mukuda99} H. Mukuda, K. Ishida, Y. Kitaoka, K. Asayama, R. Kanno, M. Takano, Phys. Rev. B \textbf{60}, 12279 (1999).
\bibitem{Gippius04} A. A. Gippius, E. N. Morozova, A. S. Moskvin, A. V. Zalessky, A. A. Bush, M. Baenitz, H. Rosner, and S.-L. Drechsler, Phys. Rev. B \textbf{70}, 020406 (2004).
\bibitem{Reuther} J.Reuther, R. Thomale S.Rachel, Phys. Rev. B \textbf{90}, 100405 (2014).
\bibitem{Kang14} B. Kang, C. Kim, E. Jo, S. Kwon, S. Lee, JMMM \textbf{360}, 1-5 (2014).
\bibitem{Narath61} A. Narath, Phys. Rev. Lett. \textbf{7}, 410 (1961).
\bibitem{note} Here, we report results for the fully relaxed crystal structure with one RuCl$_3$ layer per unit cell. Given the fact that $\alpha$-RuCl$_3$ features large amount of stacking disorder, the consideration of different stacking sequences may be important and even vital for understanding the overall magnetic order, especially in the direction perpendicular to the layers. This problem will be addressed in future studies.
\bibitem{fplo} K. Koepernik and H. Eschrig, Phys. Rev. B \textbf{59}, 1743 (1999).
\bibitem{pw92} J. P. Perdew and Y. Wang, Phys. Rev. B \textbf{45}, 13244 (1992).
\bibitem{mazin2012} I. I. Mazin, H. O. Jeschke, K. Foyevtsova, R. Valent{\'\i}, D. I. Khomskii, Phys. Rev. Lett. \textbf{109}, 197201 (2012).
\bibitem{kim08} B. J. Kim, Hosub Jin, S. J. Moon, J.-Y. Kim, B.-G. Park, C. S. Leem, Jaejun Yu, T. W. Noh, C. Kim, S.-J. Oh, J.-H. Park, V. Durairaj, G. Cao, and E. Rotenberg, Phys. Rev. Lett. \textbf{101}, 076402 (2008).
\bibitem{Kubota15}  In the final stage of the refereeing process we became aware of a paper which claims a XY-frustrated lattice for $\alpha$-RuCl$_3$. Yumi Kubota, Hidekazu Tanaka, Toshio Ono, Yasuo Narumi, Koichi Kindo, arXiv: 1503.03591 (2015).
\bibitem{Sears} Note that after the completion of this work a neutron diffraction study on single crystal was reported, which suggests a zigzag type of magnetic order below 8 K. J. A. Sears, M. Songvilay, K. W. Plumb, J. P. Clancy, Y. Qiu, and Young-June Kim, arXiv: 1411.4610 (2014).
\end{thebibliography}
\end{document}